\documentclass{article}
\usepackage{spconf,amsmath,graphicx}
\usepackage{graphicx}
\usepackage{amsmath}
\usepackage{subfigure}
\usepackage{mathrsfs}

\newcommand{\msun}{{\rm M}_{\odot}}

\title{Denoising Gravitational Waves with Enhanced Deep Recurrent Denoising Auto-Encoders}
\name{Hongyu Shen$^1$ \hspace{11pt} Daniel George$^2$ \hspace{11pt} Eliu. A. Huerta$^{2,3}$  \hspace{11pt}  Zhizhen Zhao$^{1,3}$ }
\address{
	University of Illinois at Urbana-Champaign \\
	$\ ^1$Dept. of Electrical and Computer Engineering, $\ ^2$Dept. of Astronomy,$\ ^3$NCSA }
\begin{document}
	%
	\maketitle
	\begin{abstract}
		Denoising of time domain data is a crucial task for many applications such as communication, translation, virtual assistants etc. For this task, a combination of a recurrent neural net (RNNs) with a Denoising Auto-Encoder (DAEs) has shown promising results. However, this combined model is challenged when operating with low signal-to-noise ratio (SNR) data embedded in non-Gaussian and non-stationary noise. To address this issue, we design a novel model, referred to as `Enhanced Deep Recurrent Denoising Auto-Encoder' (EDRDAE), that incorporates a signal amplifier layer, and applies curriculum learning by first denoising high SNR signals, before gradually decreasing the SNR until the signals become noise dominated. We showcase the performance of EDRDAE using time-series data that describes gravitational waves embedded in very noisy backgrounds.  In addition, we show that EDRDAE can accurately denoise signals whose topology is significantly more complex than those used for training, demonstrating that our model generalizes to new classes of gravitational waves that are beyond the scope of established denoising algorithms.
	\end{abstract}
	
	
	%
	\section{Introduction}
	\label{sec:intro}

\begin{figure*}
		\label{structure}
		\centering     
		\subfigure[DRDAE]{\label{MTDAE}\includegraphics[height=0.18\textwidth]{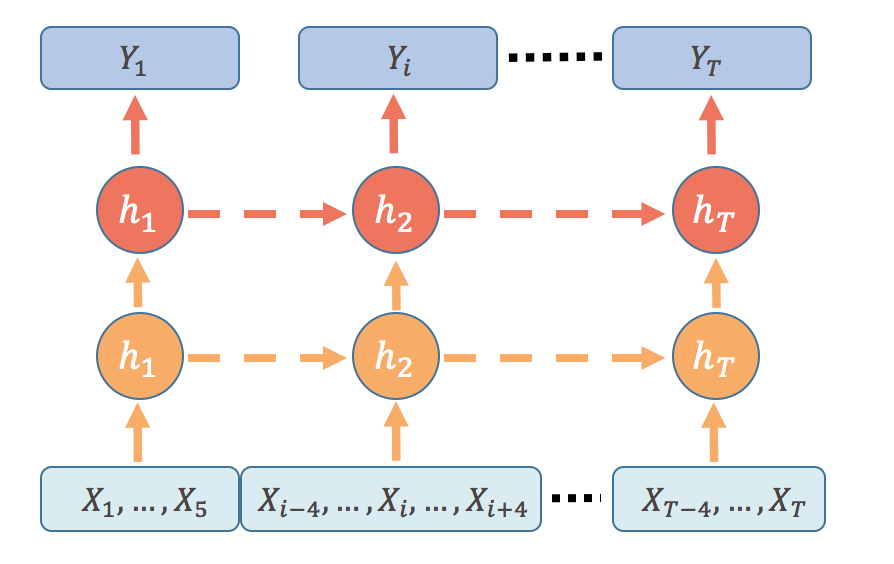}}
		\subfigure[EDRDAE]{\label{SMTDAE}\includegraphics[height=0.18\textwidth]{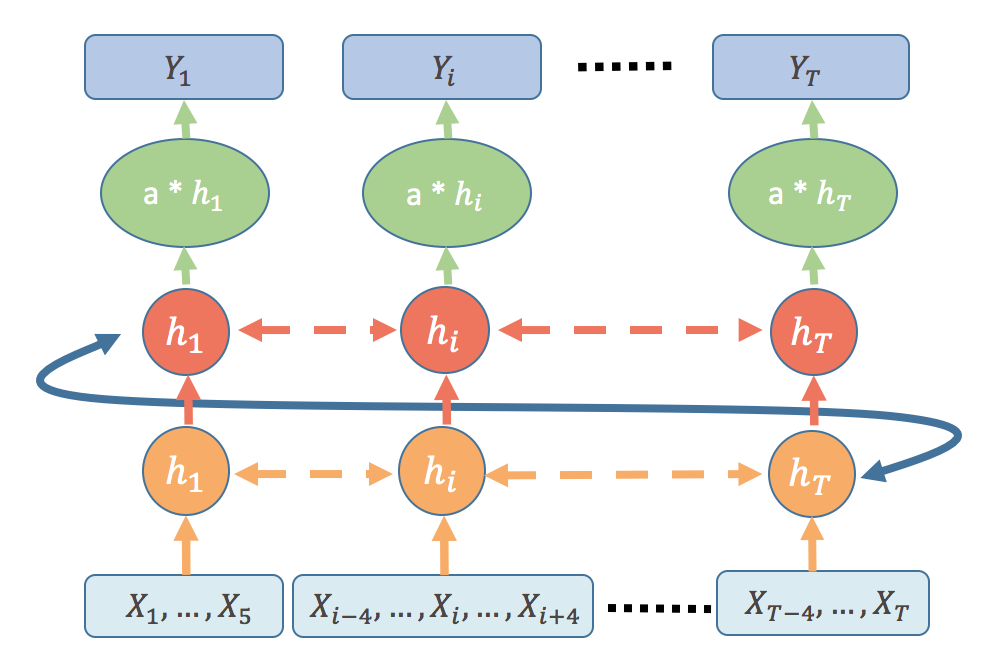}}
		\subfigure[Legend]{\label{REF}\includegraphics[height=0.18\textwidth]{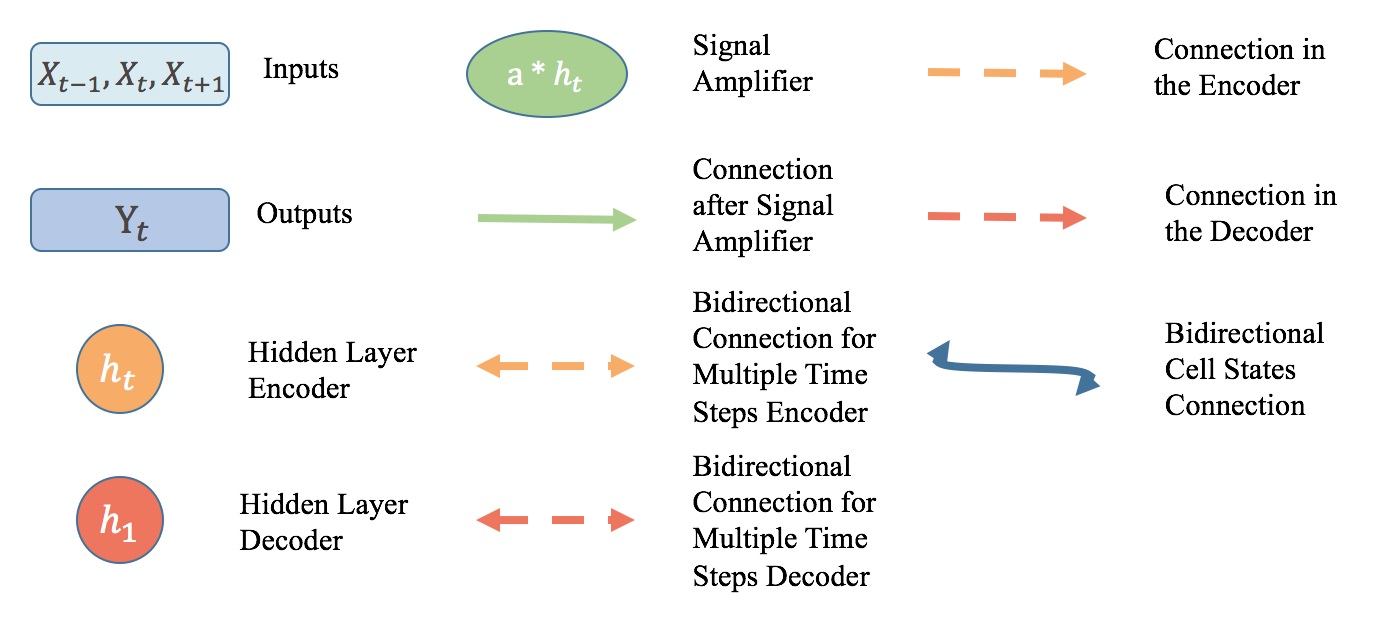}}	
		\vspace{-0.5cm}
		\caption{\subref{MTDAE} shows the DRDAE model structure. \subref{SMTDAE} represents our EDRDAE model. The nomenclature is described in~\subref{REF}. Orange circles and red circles indicate encoder and decoder part of the two models. Double arrows indicate the bidirectional LSTM connections, whereas one-directional arrows refer to conventional LSTM structure. Both DRDAE and EDRDAE has multiple timestep inputs. However, for EDRDAE, we add a cross-layer connection (blue arrows (BA)) and signal amplifier (SA) in the green ellipse. Notice that the diagrams in~\subref{MTDAE} and~\subref{SMTDAE} only display the basic structure of the models, the number of layers in the actual experiments differ and are provided in the experiment section.}
	\end{figure*}

	\indent Denoising time series data in real-time is a timely and fast-paced field of research. 
    The advent of hands-free devices for communication, translation, voice assistants, and hearing aids have triggered the development of new approaches and methods to recognize speech data in realistic, noisy environmental conditions. 
    However, denoising extremely weak time-series signals embedded in non-Gaussian and non-stationary experimental data remains an outstanding challenge. \\
    \indent The exploitation of deep learning algorithms has led to spectacular progress in signal denoising research. Denoising Auto-Encoders (DAEs) stand out among de-noising methods~\cite{Vincent:2008:ECR:1390156.1390294,DBLP:conf}. Denoising auto-encoders (DAEs) follow the architecture of AEs, a type of unsupervised deep net learns an efficient data representation~\cite{Hinton:2006:FLA:1161603.1161605,HinSal06,Bengio2006ScalingLA}, mapping an input $x$ to a hidden representation $x'$ that stores the necessary information to again reconstruct $y$ of the original input. DAEs are computationally efficient since they utilize convolutional neural networks. Convolutional structures are well-known to be effective in spectrograms. However, some time series require good algorithms in the time domain, for example, gravitational waves (GWs), due to the suboptimality of lossy non-invertible representations of the underlying data introduced by the conversion to spectrograms~\cite{dnn}, and to demands on the preserved information in time domain. Even in the time domain, the information of the front part of the data encoded into the model are almost uncorrelated with the terminal part of the data, due to the kernel structure in convolutions.  An optimal approach, Deep Recurrent Denoising Auto-Encoder (DRDAE), proposed in~\cite{DBLP:conf}, to accomplish this consists of feeding the raw time-series data to a model that takes inputs in multiple timesteps. Nevertheless, this approach is still ineffective in denoising low SNR signals, as discussed in~\cite{kumar2014fusion}. \\
    %
    \indent To address this issue, we develop (1) a new RNN DAE architecture, named Enhanced Deep Recurrent De-noising Auto-Encoder (EDRDAE) based on DRDAE, and (2) a new training strategy. We use curriculum learning~\cite{bengio2009curriculum} to ensure that data with SNRs that range from low to very high are accurately recovered. We also describe the rationale to use signal amplifiers (SA) and cross-layer state connections (blue arrows in Fig~\ref{SMTDAE}), two critical features in EDRDAE. Specifically, we demonstrate the superior performance of EDRDAE with two gravitational wave datasets, quasi-circular GWs describes two black hole mergers and eccentric GWs. We show that compared with DRDAE, EDRDAE achieves lower mean square error (MSE) and higher overlap~\cite{Canton:2014ena,Usman:2015kfa}. In addition, we show our model can generalize to different types of time series (eccentric GWs) in denoising when only one type of data (quasi-circular GWs) is used for training. Results relative to PCA and dictionary learning is also included in the experiment section. \\
    %
    \indent In this paper, we use $\mathrm{SNR_{peak}}$ to quantify the noise level, defined as $\mathrm{SNR_{peak}} = \frac{\mathrm{Maximum\ value\ of\ a\ given\ signal}}{\mathrm{Standard\ deviation\ of\ noise}}$. To facilitate direct comparisons with typical conventions in the GW community, a conversion between $\mathrm{SNR_{peak}}$ and matched-filtering SNR, $\mathrm{SNR_{MF}}$, is also provided, which is approximately $\mathrm{SNR_{MF}}=13\times\mathrm{SNR_{peak}}$.

	\section{Methods}
	\label{sec:method}
	\vspace{-3mm}
	\subsection{Model Architecture}
	\vspace{-0.35cm}
	\indent The architecture of EDRDAE (Fig~\ref{SMTDAE}) empowers this new model to better denoise time series embedded in non-Gaussian and non-stationary noise datasets. There are three major structural differences compared to DRDAE (Fig~\ref{MTDAE}). First of all, besides the multiple time step inputs, we apply bidirectional LSTMs~\cite{journals/nn/GravesS05}, rather than conventional one-way LSTMs, since the bidirectional structure will pass the information of the input data through time in two directions, forward and backward. Especially for multiple timestep inputs, where we use several neighboring time points to predict the central timestep, we found bidirectional layers to help pass information from neighboring timesteps (before and after the central timestep) to adjacent layers. Intuitively, this boosts denoising performance significantly. \\
    \indent The second difference is the introduction of the Signal Amplifier (SA) (see Fig~\ref{SMTDAE}). It is beneficial in denoising signals when the amplitude of the signal is lower than that of the background noise. This new structure is inspired by speech data, which is nearly symmetric concerning the mean-value axis (a horizontal line). SA right before the output layer assists the network in learning more evident patterns and magnifying the reconstructed values to reach the true values of the clean references.  \\
    \indent Finally, to make the training faster, we apply a cross-layer connection, which is in spirit similar to the connection in~\cite{Sutskever:2014:SSL:2969033.2969173}. It passes through both forward and backward cells in bidirectional layers. The cell states from the output timestep of the encoder are passed to the cell states of the first timestep of the decoder layer, as illustrated in Fig~\ref{SMTDAE} using blue arrows. Empirically, we observe this structure helps reconstruction in noisy environments and achieves higher accuracy compared to models that have different cell states for different layers.
	
	\begin{figure}
		\begin{center}
			\vspace{-0.5cm}
			\includegraphics[width=67mm]{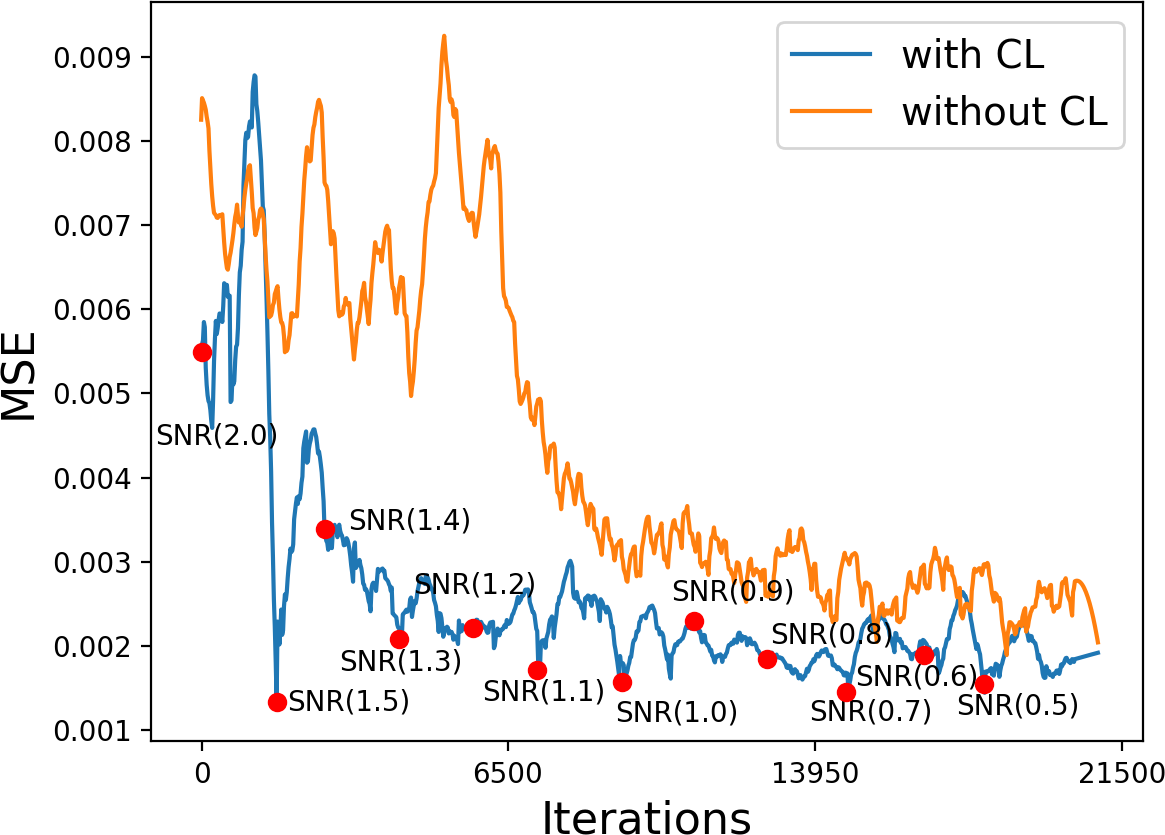}
		\end{center}
		\caption{Validation error comparison between training models with and without curriculum learning. The model with curriculum learning achieves lower MSE compared to the model without curriculum learning. The red markers indicate the change of SNRs with curriculum learning.
		}	
		\label{cl_plot}
		\vspace{-2mm}
	\end{figure}
	
	\subsection{Curriculum learning with Decreasing SNR}
	\label{dec_snr}
	
	\indent Different from the conventional image and audio denoising setups~\cite{DBLP:journals/corr/abs-1806-05229,1806.02919}. 
    for which the noise level is typically too low to obscure image background or audio utterances, the data we focus on, however, are always embedded within extreme noise. For example, raw gravitational waves. As a result, it is difficult to learn the original signal structure and remove the noise from raw data when training directly starts with very low SNRs. We found that gradually reducing SNRs during training, an idea taken from curriculum learning literature~\cite{bengio2009curriculum}, provides regularization, which allows the network to distill more accurate information of the underlying signals with larger SNRs to signals with lower SNRs. This guarantees that the information of the true signal is maintained even with extreme noise. Moreover, a decreasing SNR strategy, in combination with the previously mentioned signal amplifier (SA), ensures the reconstruction accuracy and smoothness of the data. With this approach, the model accurately denoises GWs under extremely small SNRs in testing, and it reaches lower MSE with fewer iterations compared with models trained without curriculum learning (Fig~\ref{cl_plot}).

	\begin{table}[t]
		\setlength{\tabcolsep}{1.5pt}
		\caption{CL training scheme for GW denoising. This table illustrates the change of $\mathrm{SNR_{peak}}$ happens along training for DRDAE and EDRDAE models. We start with $\mathrm{SNR_{peak}}$ 2.0 for the first 2000 iterations and then decrease $\mathrm{SNR_{peak}}$ with specified iterations in the table to the final $\mathrm{SNR_{peak}} = 0.5$. The number of iterations shown in the table where a decrease of SNR is performed is based on the loss function.}
		\label{tab1_GW}
		\vskip -0.25in
		\begin{center}
			\begin{small}
				\begin{sc}
					\begin{tabular}{cccccccccccc}
						\hline
						Iter. (k) & 2 & 3 & 4.5 & 6 & 7.5 & 9 & 10.5 &  13 & 14.5 & 16 & 17.5 \\
						$\mathrm{SNR_{peak}}$  & 1.5 & 1.4 & 1.3 &  1.2 & 1.1 & 1.0  & 0.9  & 0.8 &  0.7 & 0.6 & 0.5 \\
						\hline
					\end{tabular}
				\end{sc}
			\end{small}
		\end{center}
		\vskip -0.1in
	\end{table}

	\section{Experiment}
	\label{sec:exp} 
	
	\subsection{Gravitational Wave Datasets}
	\indent We designed experiments to illustrate the performance of DRDAE and EDRDAE on GW datasets. We use simulated GWs describe binary black hole (BBH) mergers, generated with the waveform model introduced in~\cite{bohe:2017}, which is available in LIGO's Algorithm Library~\cite{LAL}. We consider BBH systems with mass-ratios \(q\leq10\) in steps of 0.1, and with total mass \(M\in[5\msun, 75\msun]\), in steps of \(1\msun\) . The waveforms are generated with a sampling rate of 8192 Hz and whitened with the design sensitivity of LIGO~\cite{ZDHP:2010}. We consider the late inspiral, merger and ringdown evolution of BBHs, since it is representative of the BBH GW signals reported by ground-based GW detectors~\cite{DI:2016,secondBBH:2016}. 
    Specifically, we have 9861 (75\%) samples for training and 2459 (25\%) for testing. There are additional 41 eccentric waveforms with eccentricities up to 0.2 ten cycles before the black holes collide for testing the generalizability of our model. 
    For simulating non-Gaussian and non-stationary noise cases,  we take real non-Gaussian noise, 4096s from the LIGO Open Science Center (LOSC) around the LVT151012 event. \\
	
	\begin{figure}
		\begin{center}
			\includegraphics[width=63mm]{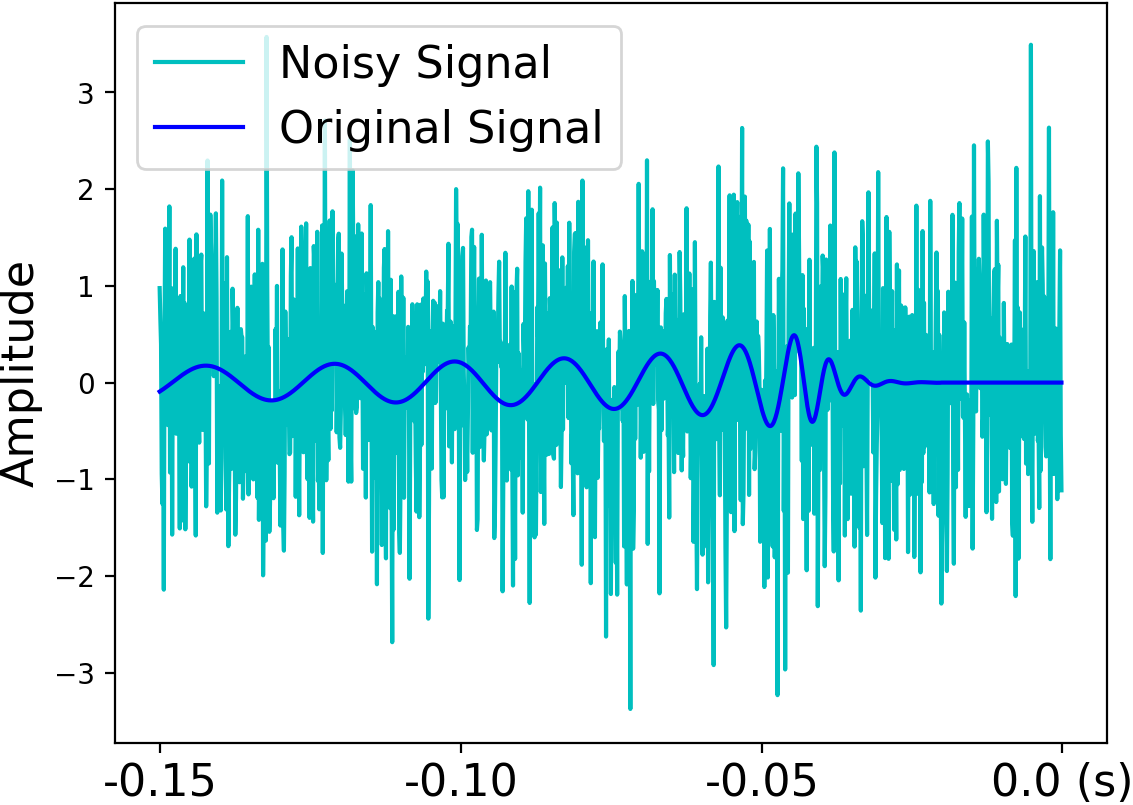}
		\end{center}
		\caption{Comparison of the noisy gravitational wave at $\mathrm{SNR_{peak}}=$ 0.5 and the clean signal.
		}	
		\label{noise_signal_plot}
	\end{figure}
	
	
	\vspace{-4mm}
	\subsection{Data Preprocessing}
	\begin{figure*}
		\centering
		\subfigure[DRDAE: quasi-circular]{\label{real_noise:img4}\includegraphics[width=53mm]{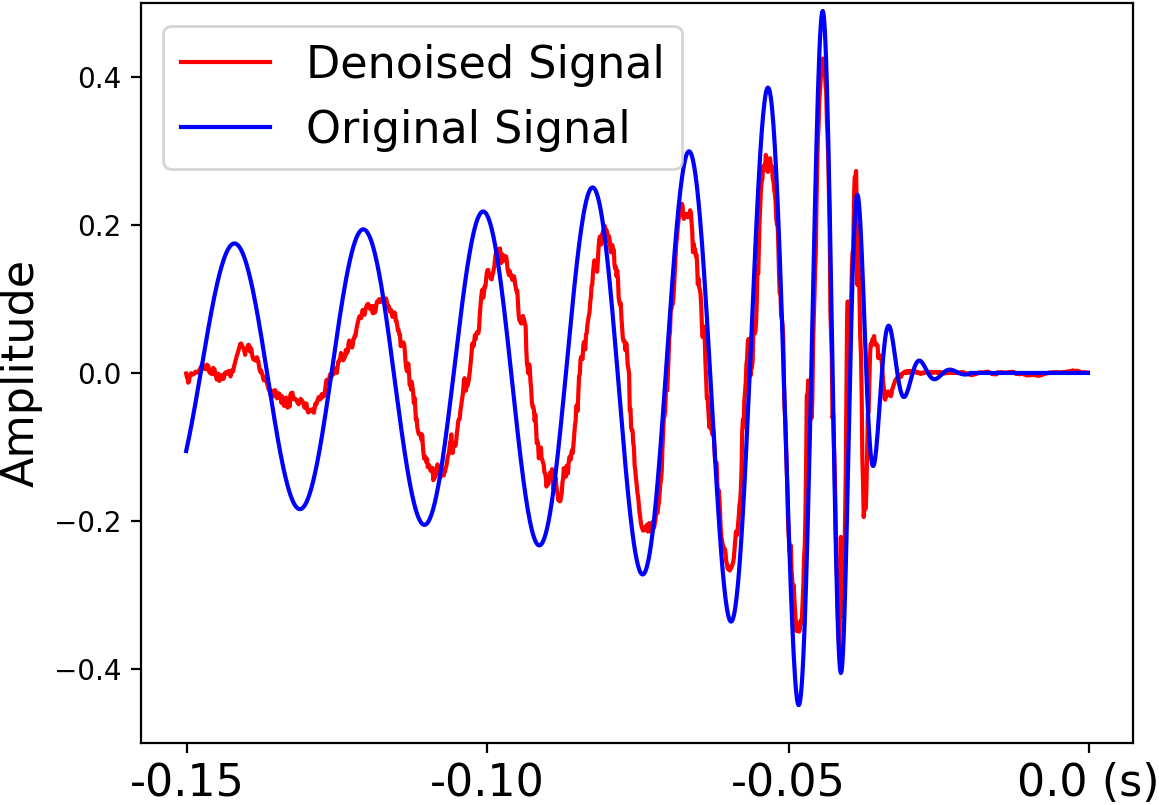}}
		\subfigure[DRDAE: pure noise]{\label{real_noise:img5}\includegraphics[width=53mm]{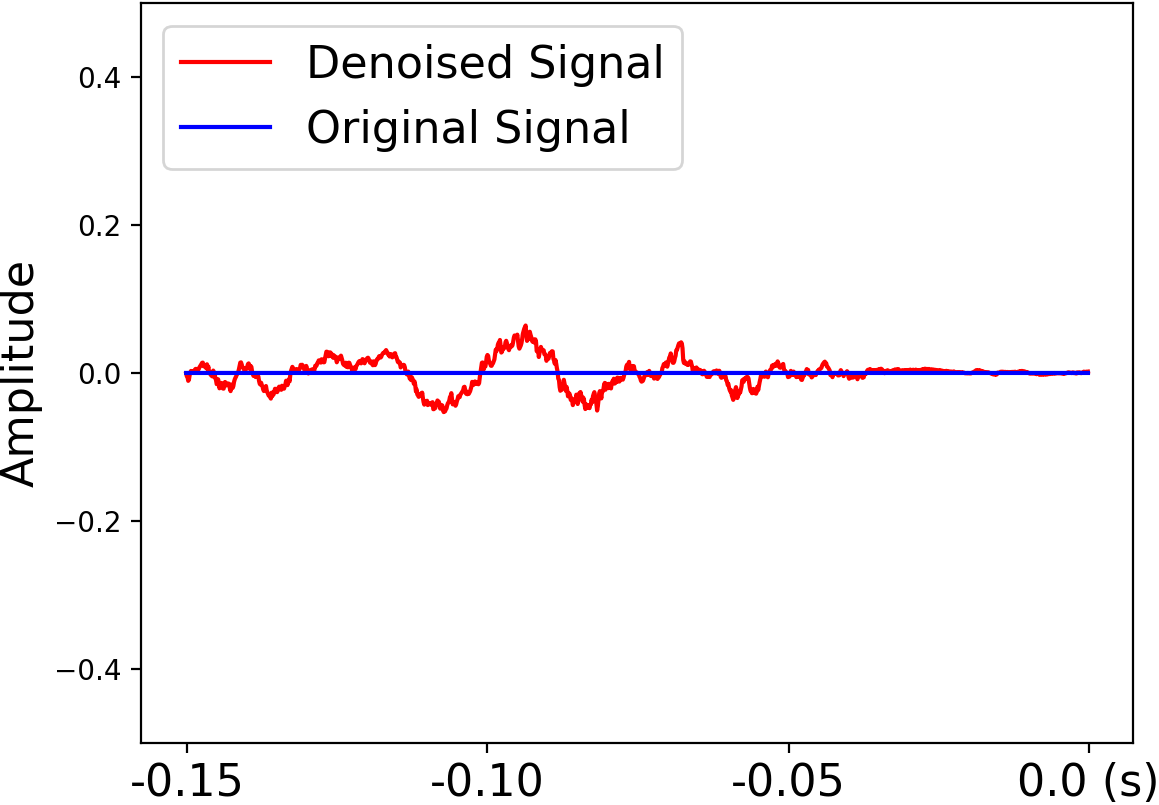}}
		\subfigure[DRDAE: eccentric]{\label{real_noise:img6}\includegraphics[width=53mm]{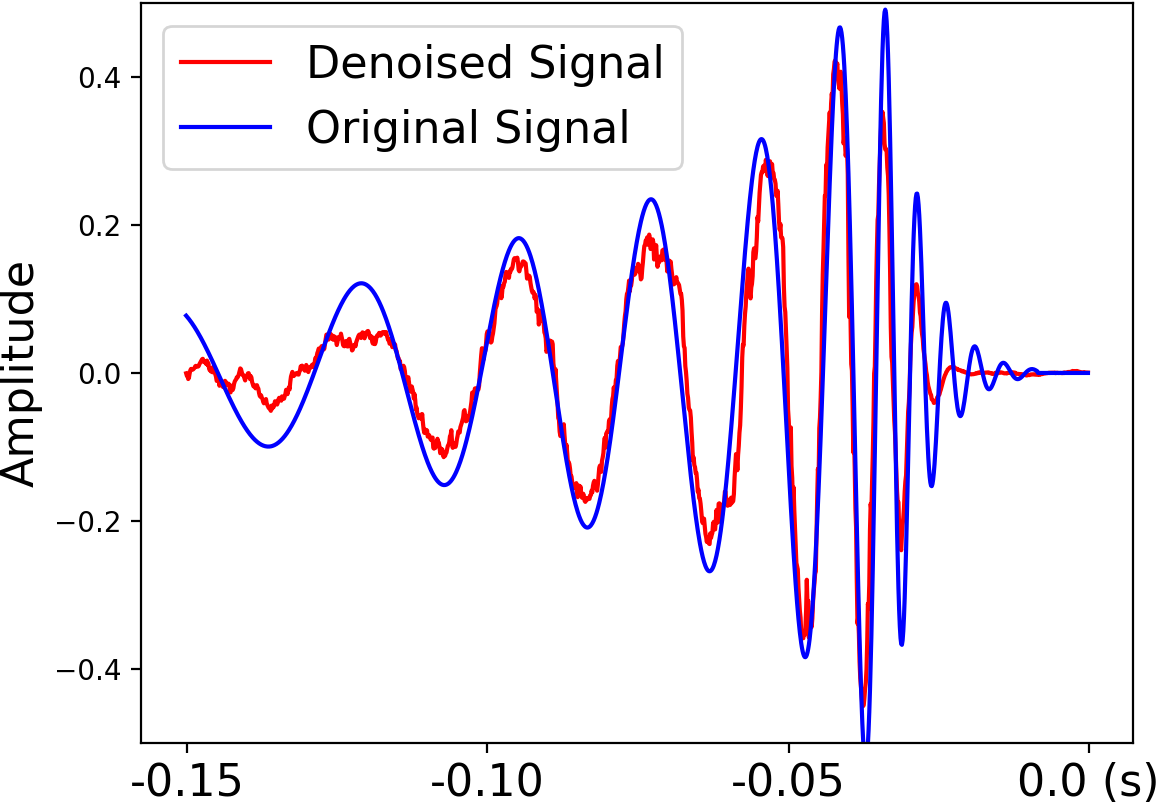}}
		\subfigure[EDRDAE: quasi-circular]{\label{real_noise:img7}\includegraphics[width=53mm]{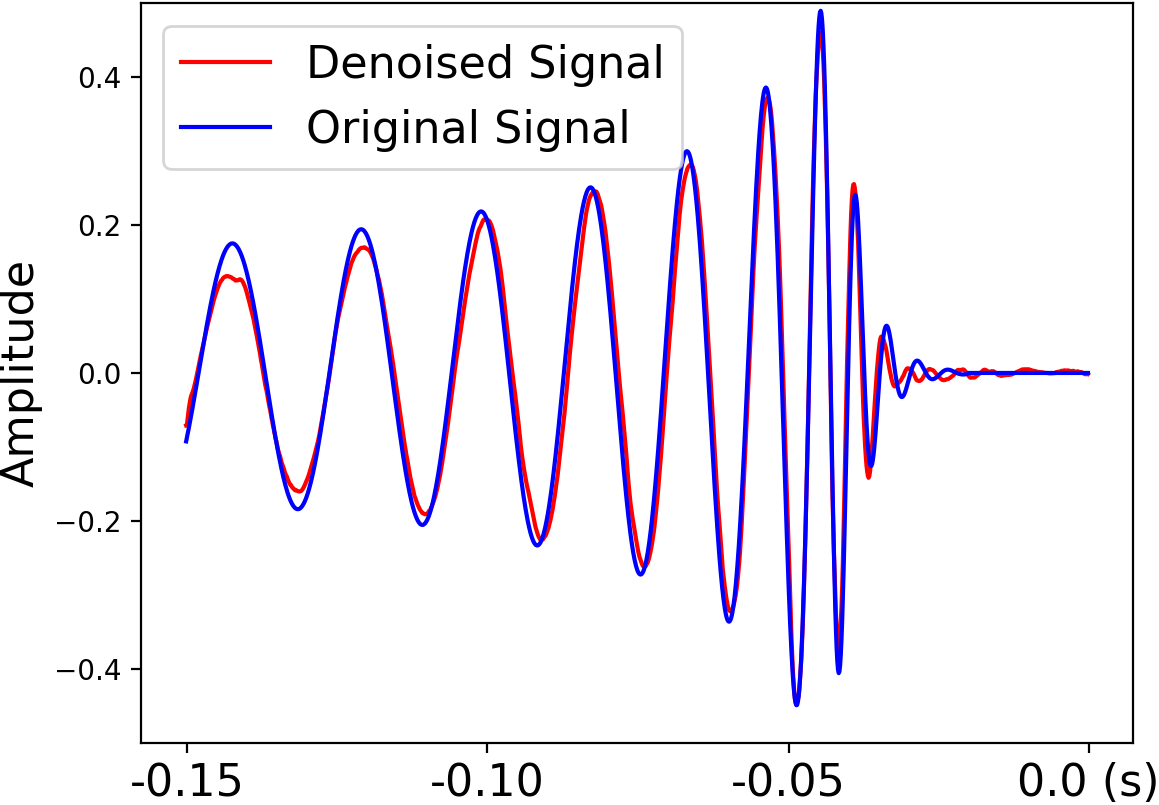}}
		\subfigure[EDRDAE: pure noise]{\label{real_noise:img8}\includegraphics[width=53mm]{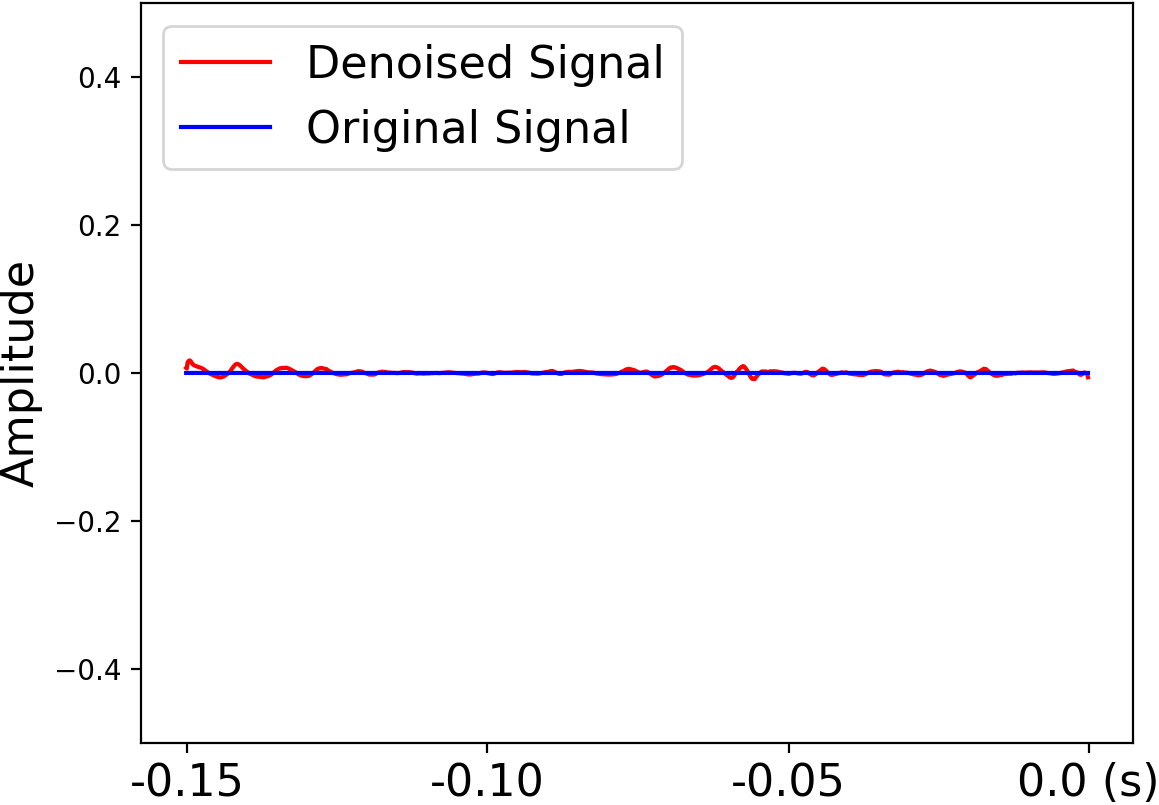}}
		\subfigure[EDRDAE: eccentric signals]{\label{real_noise:img9}\includegraphics[width=53mm]{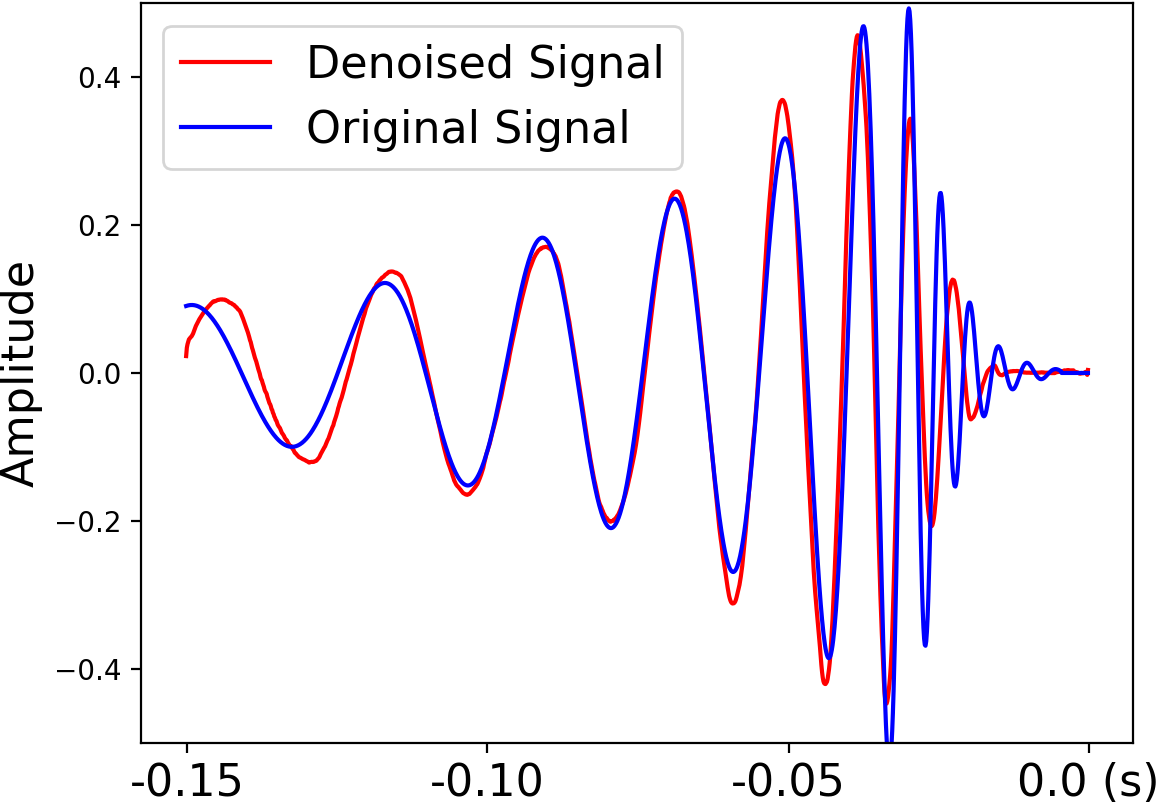}}
		\caption{Performance on GW signals contaminated by real LIGO detector noise with $\mathrm{SNR_{peak}}=0.5$. The plots include reconstructed outputs of DRDAE and EDRDAE on quasi-circular signals, pure noise input, and eccentric gravitational waves.}
		\label{real_noise_gw}
	\end{figure*}
	
	\indent The SNR of astrophysical GW sources in the LIGO detectors cannot be known prior to detection. Therefore, standardization of the datasets is a crucial step to guarantee that our model can detect GWs with a wide range of astrophysically motivated SNRs. As a result, we add additive white Gaussian noise to the clean data for each iteration during training with a predetermined $\mathrm{SNR_{peak}}$ and normalize our data to have variance 1.0. Apart from the timestep augmentation~\cite{DBLP:conf}, we also add random shifts to the training data, to make the model more resilient to variations in the location of the signal. For every input signal, we randomly generate an integer between 0 and 200 as shift length for left and right shifts. The length is 0\% to 15\% proportional to the total signal length. Zero padding is performed when necessary. 
	
	\begin{table}[t]
		\setlength{\tabcolsep}{1.5pt}
		\caption{Comparisons of MSE and overlap across different approaches. $\cdot/\cdot$ refers to "metric for quasi-circular" / "metric for eccentric". Here "DL" refers to dictionary learning. "WT" refers to wavelet thresholding.}
		\label{tab2_metric}
		\vskip -0.25in
		\begin{center}
			\begin{small}
				\begin{sc}
					\begin{tabular}{cccccc}
						\hline
						Model & PCA & DL & WT & DRDAE & EDRDAE \\
						\hline
						MSE & .033/.034 & .108/.107& .017/.018  &.008/.004&  .001/.005 \\
						Overlap & .644/.639 & .466/.476& .684/.671& .906/.955 &  .994/.985  \\  
						\hline
					\end{tabular}
				\end{sc}
			\end{small}
		\end{center}
		\vskip -0.1in
	\end{table}
	
	\subsection{Training and Evaluation}
	\label{trainAndeva}
	
	\indent All trainings are performed on NVIDIA Tesla V100 GPUs using TensorFlow~\cite{1603.04467}. We train the models with batch size 30. To ensure a fair model comparison, each encoder and decoder has four layers for DRDAE, with a one-way LSTM cell that has 64 channels for each layer. For EDRDAE, we have one encoder layer followed by two decoder layers with bidirectional LSTM cells. With the defined structures, DRDAE has 250,177 parameters, and EDRDAE has 235,650 parameters. We tuned the optimizers (ADAM, RMSprop, and SGD), learning rate (0.0001 to 0.001), batch size (10 to 50), number of layers (2 to 4) and the size of input multiple time steps (5 to 20). Only results of the best model (Adam optimizer, 0.0001 learning rate, batch size 30, 4-layer model 9 timesteps for input) displayed in Fig~\ref{real_noise_gw}. We use the default ADAM setup described in~\cite{journals/corr/KingmaB14}. To denoise signals with extremely low SNR, for example, $\mathrm{SNR_{peak}}=0.5$ (Fig~\ref{noise_signal_plot}), our training starts with $\mathrm{SNR_{peak}}=2.0$. We gradually decrease the SNR at every subsequent training step until we reach a $\mathrm{SNR_{peak}}=0.5$ (Shown in Table~\ref{tab1_GW} and Fig~\ref{cl_plot}). We only input BBH GWs and additive Gaussian noise for training. Real LIGO noise and both BBH GWs and eccentric GWs are used for testing. In general, the models become stable after training for about 20,000 iterations. We summarize results after the final iteration of training in Fig~\ref{real_noise_gw}. Two metrics are used for model evaluation for GWs. Mean Squared Error (MSE) is a conventional metric for GW evaluation measuring the reconstruction quality. We also evaluate a metric known as overlap, which measures how similar any two GWs are. Overlap values equal to unity indicate perfect agreement between any two given signals. For GW detection, LIGO typically requires a minimum overlap value 0.97~\cite{Canton:2014ena,Usman:2015kfa}. Measurements are concluded in Table~\ref{tab2_metric} with benchmarks that include PCA with Wiener filtering and dictionary learning optimized by coordinate descent, as well as wavelet universal thresholding with 10 levels~\cite{Donoho94idealdenoising}. An ablation study for the SA, cross-layer connections and decreasing SNRs is also provided in Table~\ref{tab3_ablation}.
	
	\begin{table}[t]
		\setlength{\tabcolsep}{1.5pt}
		\caption{Ablation study for major parts in EDRDAE}
		\label{tab3_ablation}
		
		\begin{center}
			\begin{small}
				\begin{sc}
					\begin{tabular}{cccc}
						\hline
						Missing Parts & W/O SA & W/O BA & W/O CL \\
						\hline
						MSE & .017/.002 & .007/.003 & .010/.001 \\
						Overlap & .769/.979 & .941/.974 & .941/.960  \\  
						\hline
					\end{tabular}
				\end{sc}
			\end{small}
		\end{center}
		\vskip -.2in
	\end{table}
	
	
	Results on DRDAE and EDRDAE with real LIGO noise are in Fig~\ref{real_noise_gw}. We compare the denoising results for a quasi-circular GW in Fig~\ref{real_noise:img4} and Fig~\ref{real_noise:img7}. 
    To prevent spurious detection, which is crucial in GW detection and denoising, we also test the model when the input is only pure noise. 
    Notice that our model is also able to prevent fake wave generation from pure noise (see Fig.~\ref{real_noise:img8}), whereas the output from DRDAE still contains wave-like structure with noticeable amplitude (see Fig.~\ref{real_noise:img5}). Additionally, we also test the resilience of the model to denoise signals that are not used during training, i.e., eccentric GWs.  Currently, there is no signal-processing algorithm adequate for the detection or denoising of eccentric GWs due to their complex topology~\cite{2017PhRvD..95b4038H}. We chose eccentric GWs to showcase the robustness of our algorithm to denoise signals that are significantly different from the training data. The sample waveforms are presented in Fig~\ref{real_noise:img6} and Fig~\ref{real_noise:img9}. It is clear our model achieves better reconstruction in both quasi-circular GWs and eccentric GWs with lower MSEs and higher overlaps (Table~\ref{tab2_metric}).  

	\section{Conclusion}
	\label{sec:conclusion}	
	\indent In this paper, we proposed a new deep recurrent denoising autoencoder to denoise gravitational wave signals contaminated by an extremely high level of noise often encountered in realistic detection scenarios. By introducing additional structures to the model (cross-layer connection, signal amplifier), and by adopting a training approach that gradually reduces the SNRs of the training samples, we show that our model outperforms DRDAE and other tested popular denoising algorithms (PCA, dictionary learning and wavelet thresholding) for GW denoising. It is also noteworthy that although our denoising auto-encoder was only trained with quasi-circular GWs contaminated with additive white Gaussian noise, it is able to handle both quasi-circular GWs with different mass ratios and eccentric GWs embedded in real LIGO noise. Therefore the proposed method has great generalization performance. 

	\clearpage
	
	\bibliographystyle{IEEEbib}
	\bibliography{references}
	
\end{document}